\documentclass[conference, a4paper]{IEEEtran}
\IEEEoverridecommandlockouts
\usepackage{cite}
\usepackage{amsmath,amssymb,amsfonts}
\usepackage{algorithmic}
\usepackage{graphicx}
\usepackage{textcomp}
\usepackage{subfig}
\usepackage{verbatim}
\def\BibTeX{{\rm B\kern-.05em{\sc i\kern-.025em b}\kern-.08em
    T\kern-.1667em\lower.7ex\hbox{E}\kern-.125emX}}
\begin{document}
\title{Grayscale-Based Image Encryption Considering Color Sub-sampling Operation
for Encryption-then-Compression Systems}

\author{\IEEEauthorblockN{Warit Sirichotedumrong, Tatsuya Chuman and Hitoshi
Kiya } \IEEEauthorblockA{Tokyo Metropolitan University, Asahigaoka, Hino-shi,
Tokyo, 191-0065, Japan}
\thanks{This work was partially supported by Grant-in-Aid for Scientific
Research(B), No.17H03267, from the Japan Society for the Promotion Science.} }
\maketitle

\begin{abstract}
A new grayscale-based block scrambling image encryption scheme is presented to
enhance the security of Encryption-then-Compression (EtC) systems, which are
used to securely transmit images through an untrusted channel provider. The
proposed scheme enables the use of a smaller block size and a larger number of
blocks than the conventional scheme. Images encrypted using the proposed scheme
include less color information due to the use of grayscale images even when the
original image has three color channels. These features enhance security against
various attacks, such as jigsaw puzzle solver and brute-force attacks. Moreover,
it allows the use of color sub-sampling, which can improve the compression
performance, although the encrypted images have no color information. In an
experiment, encrypted images were uploaded to and then downloaded from Facebook
and Twitter, and the results demonstrated that the proposed scheme is effective
for EtC systems, while maintaining a high compression performance.
\end{abstract}

\begin{IEEEkeywords}
Compression, encryption, EtC systems
\end{IEEEkeywords}

\section{Introduction}
\label{sec:intro}
The use of images and video sequences has greatly increased because of rapid
growth of the Internet and multimedia systems. A lot of studies on secure,
efficient and flexible communications have been
reported\cite{huang2014survey,lagendijk2013encrypted,zhou2014designing}.
For securing multimedia data, full encryption with provable security (like RSA,
AES, etc) is the most secure options. However, many multimedia applications have
been seeking a trade-off in security to enable other requirements, e.g., low
processing demands, retaining bitstream compliance, and signal processing in the
encrypted domain.

Encryption-then-Compression (EtC) systems with JPEG
compression\cite{Johnson_2004,Liu_2010,Hu_2014,zhou2014designing} have been
proposed to be applied to Social Network Services (SNS) and Cloud Photo Storage
Services (CPSS). However, the color-based image encryption schemes for EtC
systems\cite{watanabe2015encryption,kurihara2015encryption,KURIHARA2015,Kuri_2017}
cannot provide the robustness against color sub-sampling used for JPEG
compression because an encrypted image is a full-color image. In order to solve
this issue, the grayscale-based images encryption has been
proposed\cite{WARIT2018ICME,WARIT2018ITCCSCC} to encrypt a full-color image as a
grayscale-based image. Even if the grayscale-based image
encryption\cite{WARIT2018ICME} can avoid the effect of color sub-sampling, it
is impossible to consider color sub-sampling operation because the grayscale-based image is generated from RGB components. Moreover,
compared to the color-based image
encryption\cite{watanabe2015encryption,kurihara2015encryption,KURIHARA2015,Kuri_2017},
the compression performance is strongly degraded. According to
\cite{WARIT2018ITCCSCC}, the grayscale-based image encryption generated from
YCbCr components and the quantization table for grayscale-based images have been
proposed to provide the better compression performance. However, the color sub-sampling operation has not been considered.

This paper discusses and considers the color sub-sampling operation for
grayscale-based image encryption. Instead of generating the grayscale-based
image from RGB components, a full-color image in RGB color space is firstly
transformed to YCbCr color space. Hence, color sub-sampling operation can be
performed to generate grayscale-based images. Moreover, we describe scenario and requirements that image encryption have to satisfy. The enhancements of
compression performance and robustness against color sub-sampling are evaluated
in terms of Rate-Distortion (R-D) curves.

\section{Scenario and Requirements}
\label{sec:scenario}

\begin{figure}[t]
\centering
\includegraphics[width =8.7 cm]{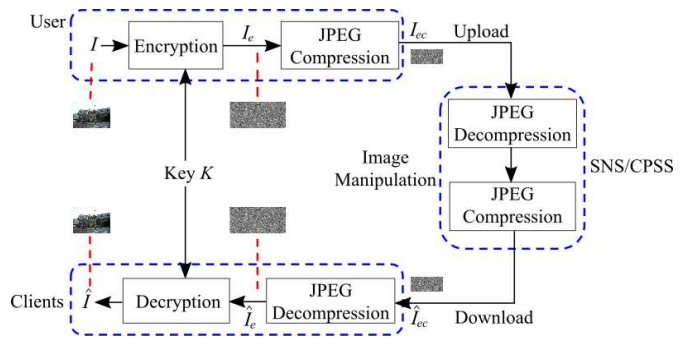}
\caption{EtC system}
\label{fig:etc}
\end{figure}

According to the image manipulation on Social Network Services
(SNS) and Cloud Photo Storage Services (CPSS)\cite{CHUMAN2017APSIPA}, almost all
providers manipulate every uploaded image as illustrated in Fig.\,\ref{fig:etc}.
Because of such scenarios, the grayscale-based image encryption schemes for EtC
system have been proposed\cite{WARIT2018ICME,WARIT2018ITCCSCC} as the extension
of the color-based EtC
systems\cite{watanabe2015encryption,kurihara2015encryption,KURIHARA2015,Kuri_2017}.

There are three requirements that image encryption schemes have to satisfy:
compression performance, security level, and robustness against image
manipulation.

\subsubsection{Compression Performance}
\label{ssec:compression}

\begin{figure}[t]
\centering
\includegraphics[width =8.7 cm]{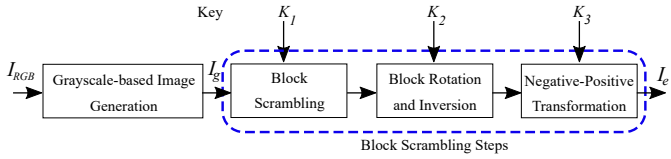}
\caption{Conventional grayscale-based block scrambling image encryption}
\label{fig:conv_step}
\end{figure}

In order to apply an image encryption scheme to SNS and CPSS, it is necessary
for encrypted JPEG images to have almost the same compression performance as the
non-encrypted ones. The color-based encryption
scheme\cite{watanabe2015encryption,kurihara2015encryption,KURIHARA2015,Kuri_2017}
can provide almost the same compression performance as the non-encrypted JPEG
images. However, it cannot be achieved by the
conventional grayscale-based image encryption\cite{WARIT2018ICME} because a
grayscale-based image is generated from RGB components as shown in Fig.\,\ref{fig:conv_step}

\subsubsection{Security Level}
\label{ssec:security}

In this paper, we consider security against brute-force attack and jigsaw puzzle
solver attacks as ciphertext-only attacks.
It has been confirmed that the key space of block scrambling-based image
encryption for EtC systems is huge enough against brute-force
attack\cite{KURIHARA2015} and has the robustness against jigsaw puzzled solver
attacks\cite{Son_2016_CVPR,Cho_2010_CVPR,CHUMAN2017ICASSP,CHUMAN2017ICME,WARIT2018ICME}.

This paper considers the extended jigsaw puzzle
solver\cite{CHUMAN2017ICASSP,CHUMAN2017ICME} as ciphertext-only attacks. There
are three metrics using for evaluating
the robustness against jigsaw puzzle
solver attacks\cite{Gallagher_2012_CVPR,Cho_2010_CVPR} which are described as
follows:
\begin{itemize}
	\item {\bf Direct comparison ($D_c$)} is the ratio between the number of pieces
	which are placed in the correct position and the total number of pieces.
	\item {\bf Neighbor comparison ($N_c$)} expresses the ratio of the number of pieces that are joined with the correct pattern and the total number of pieces.
	\item {\bf Largest components ($L_c$)} refers to the ratio between the number
	of the largest joined blocks that are correctly adjacent and the number of pieces.
\end{itemize}
Note that $D_c$, $N_c$, $L_c$ $\in [0,1]$ and a larger value means a higher
compatibility.

As images encrypted using the grayscale-based image encryption contain only one
color channel\cite{WARIT2018ICME}, the smallest block size ($B_x \times B_y$) of
the grayscale-based image encryption is $8 \times 8$. Moreover, since
the block size is smaller, and the number of blocks is larger than the
color-based encryption scheme. As a result, grayscale-based encrypted images
have stronger security and robustness against jigsaw puzzled solver
attacks\cite{CHUMAN2017ICME} than those with the color-based one.

\subsubsection{Robustness against Image Manipulation}
\label{sssec:manipulation}

It is known that almost all SNS and CPSS providers manipulate every uploaded
image when it satisfies their conditions\cite{CHUMAN2017APSIPA}. Uploaded JPEG
images are decompressed and sequentially recompressed with new compression
parameters based on their algorithms. In recompression, as the color
sub-sampling is usually carried out, the image encryption which has robustness
against color sub-sampling is required. However, this requirement cannot be
achieved by the color-based encryption
scheme\cite{watanabe2015encryption,kurihara2015encryption,KURIHARA2015,Kuri_2017}
while the conventional grayscale-based image
encryption\cite{WARIT2018ICME,WARIT2018ITCCSCC} has been proposed to avoid the
effect of color sub-sampling.

\begin{figure}[t]
\centering
\includegraphics[width =8.7 cm]{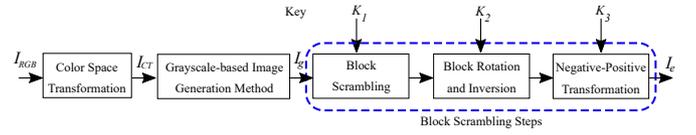}
\caption{Proposed grayscale-based block scrambling image encryption}
\label{fig:step}
\end{figure}
\section{Proposed Grayscale-based Image Encryption}
\label{sec:proposed}

This section describes an encryption procedure of the proposed grayscale-based
image encryption and how the color sub-sampling operation is considered with the
encryption scheme. Finally, the quantization table for grayscale-based images is
discussed.

\subsection{Encryption Procedure}
\label{ssec:encryption}

Let us consider a full-color image ($I_{RGB}$) with $M \times N$ pixels. To
encrypt $I_{RGB}$, the following six steps are carried out as follows (See
Fig.\,\ref{fig:step}).
\begin{itemize}
  \item [Step1:] $I_{RGB}$ are transformed into the full-color image in YCbCr
  color space ($I_{YCbCr}$), so that $I_{CT}=I_{YCbCr}$.
  \item [Step2:] Luminance ($i_Y$) and chrominance ($i_{Cb}$ and
  $i_{Cr}$) are concatenated vertically or horizontally to generate the
  grayscale-based image ($I_g$) with $3(M \times N)$ pixels.
  \item [Step3:] $I_g$ with $M_g \times N_g$ pixels is divided into
  non-overlapping blocks each with $B_x \times B_y$. The number of divided
  blocks, $N_b$, is expressed by
\begin{equation}
N_b = \lfloor \frac{M_g}{B_x} \rfloor \times \lfloor \frac{N_g}{B_y} \rfloor
\end{equation}
where $\lfloor \cdot \rfloor$ is the floor function that rounds down to the
nearest integer.
	\item [Step4:] Randomly permute the divided blocks based on a random integer
	which is generated by a secret key $K_1$.
	\item [Step5:] Rotate and invert each divided block randomly based on a random
	integer generate by a secret key $K_2$.
	\item [Step6:] Perform the negative-positive transformation to each divided
	block using a random binary integer generated by a secret key $K_3$.
	A transformed pixel of $i$th block is represented by $p'$ and can be expressed
	as 
	\begin{equation}
	p'=
	\left\{
	\begin{array}{ll}
	p & (r(i)=0) \\
	p \oplus (2^L-1) & (r(i)=1)
	\end{array}
	\right.
	\end{equation}
	where $r(i)$ is a random binary integer generated by $K_3$ and $p$ is the pixel
	value of an original image with $L$ bits per pixel.
\end{itemize}

\subsection{Color sub-sampling for Grayscale-based Images}
\label{sec:sub-sampling}
\begin{figure}[t]
\centering
\includegraphics[width =8.7 cm]{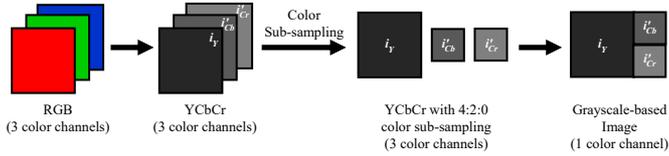}
\caption{Grayscale-based image generation method}
\label{fig:gray_gen}
\end{figure}

\begin{figure}[t]
\captionsetup[subfigure]{justification=centering}
\centering
\subfloat[Grayscale-based image with 4:2:0 color
sub-sampling]{\includegraphics[clip, height=2cm]{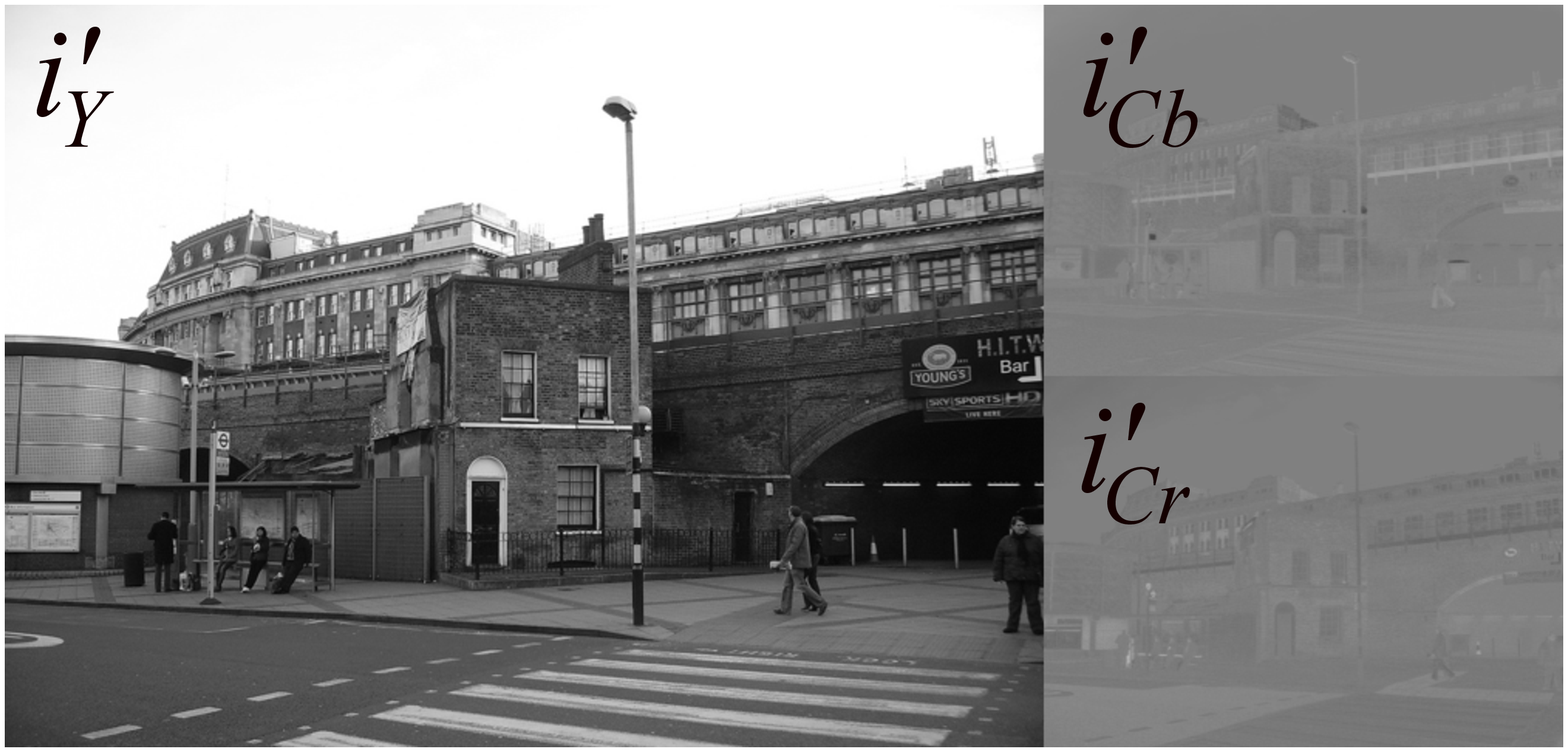}}
\hfil
\subfloat[Image encrypted by grayscale-based image
encryption considering 4:2:0 color sub-sampling]{\includegraphics[clip,
height=2cm]{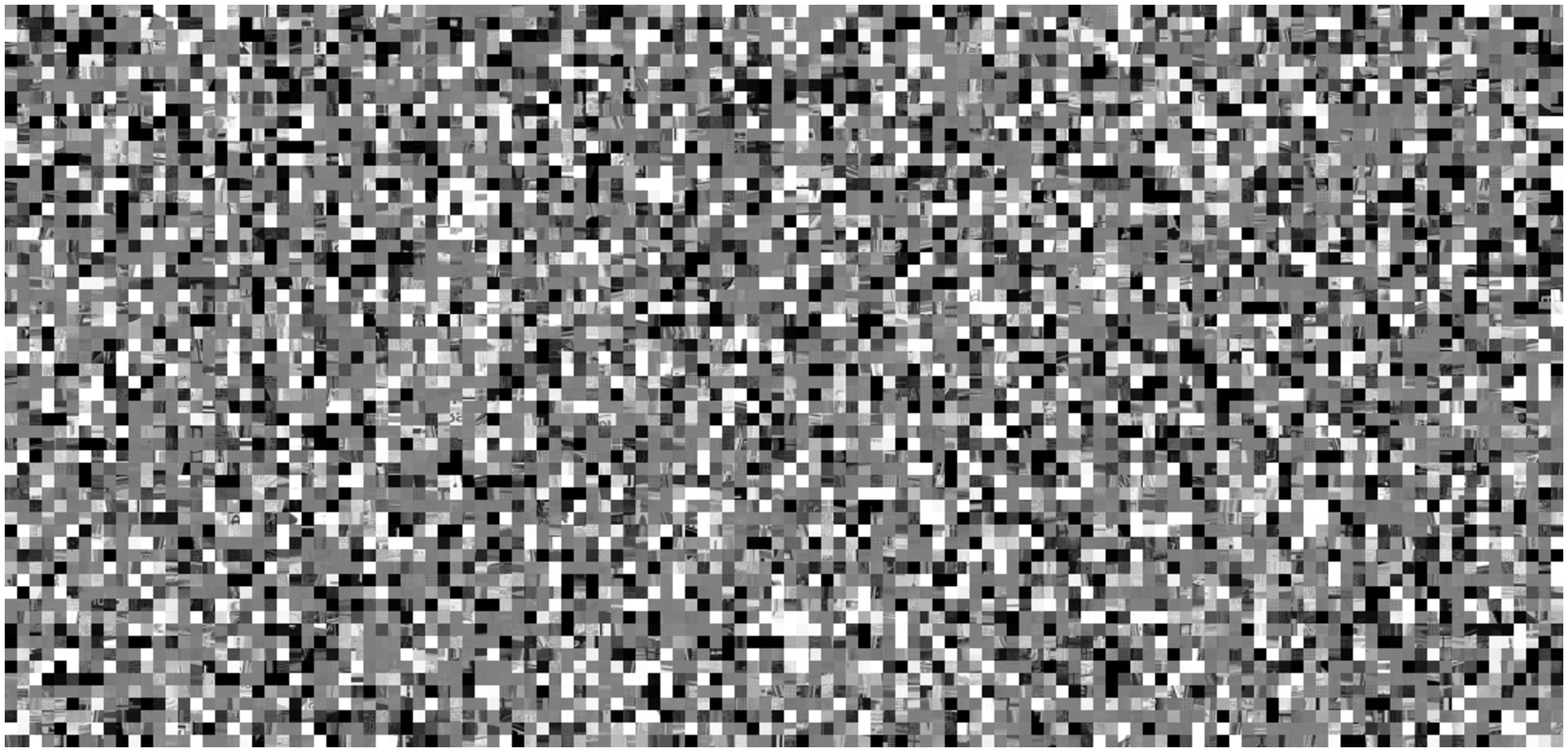}}

\caption{Example images}

\label{fig:ex_img}
\end{figure}

As previously described in Section\,\ref{ssec:encryption} that $I_{RGB}$ is
firstly transformed to $I_{YCbCr}$, this paper considers the color
sub-sampling operation for the grayscale-based image encryption. Since human eyes are more sensitive to $i_Y$ than $i_{Cb}$ and $i_{Cr}$, we
downsample $i_{Cb}$ and $i_{Cr}$ using 4:2:0 color sub-sampling operation
provided by IJG software\cite{JPEGLIB} as shown in Fig.\,\ref{fig:gray_gen}. The
sub-sampled chrominance components are represented by $i'_{Cb}$ and $i'_{Cr}$. Eventually, $i_Y$, $i'_{Cb}$, and
$i'_{Cr}$ are combined to produce $I_g$ with $\frac{3}{2}(M \times N)$ pixels.
The example of $I_g$ with 4:2:0 color sub-sampling is shown in
Fig.\,\ref{fig:ex_img}(a). As $I_g$ has only one color channel, the
color sub-sampling operation is not carried out. Thus, the proposed scheme can
provide the robustness against color sub-sampling and better compression
performance.

\subsection{Quantization Table for Grayscale-based Images}
\label{ssec:gtable}

JPEG softwares, such as Independent JPEG Group (IJG) software \cite{JPEGLIB},
generally utilize two default quantization tables to quantize $i_Y$, $i_{Cb}$,
and $i_{Cr}$ of $I_{YCbCr}$ where $i_Y$ is quantized by the luminance
quantization table (Y-table), and the chrominance quantization table
(CbCr-table) is employed to quantize $i_{Cb}$ and $i_{Cr}$. However, users are
allowed to use other tables rather than the default ones. The image-dependent
quantization table has been proposed to minimize the distortion of quantization
process of each block\cite{YANG2009}. However, since $I_g$ is generated from $i_Y$, $i_{Cb}$,
and $i_{Cr}$, those tables are not designed for $I_g$. Therefore, the
quantization table called G-table has been proposed to improve the compression
performance of $I_g$\cite{WARIT2018ITCCSCC}.

In JPEG compression, all pixel values in each block of $I_g$ are mapped from
$[0,255]$ to $[-127,128]$ by subtracting 128, then each block is transformed
using Discrete Cosine Transform (DCT) to obtain DCT coefficients.

The DCT coefficients are employed to generating G-table. Let $D_n(i,j)$ be the DCT coefficient
of the $n^{th}$ block at the position $(i,j)$ where $1\leq i\leq 8$ and $1\leq
j\leq 8$. Considering every block of $I_g$, the Euclidean distance between the
origin $O$ and $D_n(i,j)$ is measured, and the arithmetic mean of the
distance is expressed by

\begin{equation}
\label{eq:mean_in_img}
c(i,j)=\frac{1}{N_b}\sum_{n=1}^{N_b}|D_n(i,j)-O|
\vspace{-0.1cm}
\end{equation}
where $I_g$ consists of $N_b$ blocks.

As a set of grayscale-based images which consists of $R$ images is utilized to
determine G-table, we define $c_n(i,j)$ as $c(i,j)$ of the $n^{th}$ image and
calculate the average of every $c(i,j)$ from $R$ grayscale-based images. The
average $\bar{c}(i,j)$ is calculated as follow.

\begin{equation}
\label{eq:mean_in_dataset}
\bar{c}(i,j)=\frac{1}{R}\sum_{n=1}^{R}c_n(i,j)
\end{equation}

To obtain G-table, $q(i,j)$ represents the quantization step size
at $(i,j)$ and is derived from the ratio between $\bar{c}(1,1)$ and
$\bar{c}(i,j)$. The step size can be calculated by

\begin{equation}
\label{eq:q_tbl}
q(i,j)=\bigl\lceil \frac{\bar{c}(1,1)}{\bar{c}(i,j)} \bigr\rceil+\epsilon
\end{equation}
where $\epsilon$ is set to 16 for adjusting the Y-table step size at $(1,1)$
as for IJG software\cite{JPEGLIB}.

\section{Experiments}
\label{sec:experiment}

\subsection{Experimental Set-up}
\label{ssec:setup}

To evaluate the performance of the grayscale-based image
encryption considering color sub-sampling operation, this paper utilizes two
datasets as below.

\begin{itemize}
  \item[(a)] 20 images from MIT dataset ($672 \times 480$)\cite{Cho_2010_CVPR}
  \item[(b)] 1338 images from Uncompressed Color Image Database
  (UCID)\cite{ucid}
\end{itemize}
\par
All images in Dataset (a) were encrypted using the proposed scheme and
conventional one with $B_x=B_y=8$. Then, all encrypted images were compressed
with specific quality factors, $Q_{f} \in [70,100]$, using the JPEG standard from IJG
software\cite{JPEGLIB}.

\begin{figure}[t]
\centering
\includegraphics[width =5 cm]{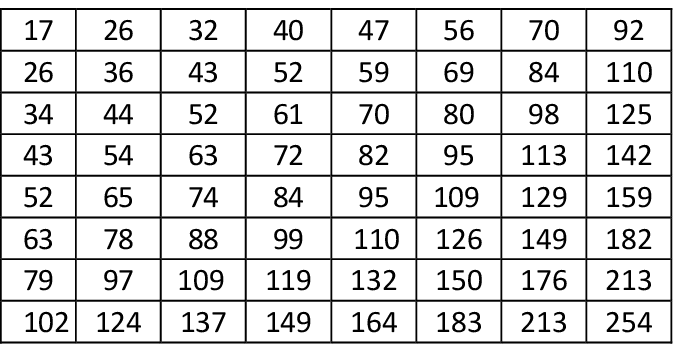}
\caption{G-table for grayscale-based images}
\label{fig:g-table}
\end{figure}

All images in dataset (b) were compressed using IJG software\cite{JPEGLIB} to obtain compressed
grayscale-based images. Note that DCT coefficients are extracted
during this JPEG compression. According to the procedures in
section\,\ref{ssec:gtable}, G-table was designed by using the DCT coefficients
whereas $N_b=4608$, $R=1338$, and $\epsilon=16$. As a result, G-table is shown
in Fig.\,\ref{fig:g-table}.
 We conduct two experiments: without color sub-sampling (4:4:4) and with color
 sub-sampling (4:2:0). All JPEG images were decompressed and measure Peak-Signal-to-Noise Ratio (PSNR), respectively.

\subsection{Results and Discussions}
\label{ssec:result}

We evaluated the compression performance and robustness against color
sub-sampling of the proposed scheme based on Rate-Distortion (R-D) curves which
are the relation between the arithmetic mean PSNR of the images and bits per
pixel ($bpp$) of JPEG images. The proposed grayscale-based image encryption was
compared with the non-encrypted images.
\subsubsection{Compression Performance of Uploaded Images}
\label{sssec:performance_up}
\begin{figure}[t]
\centering
\includegraphics[width =8. cm]{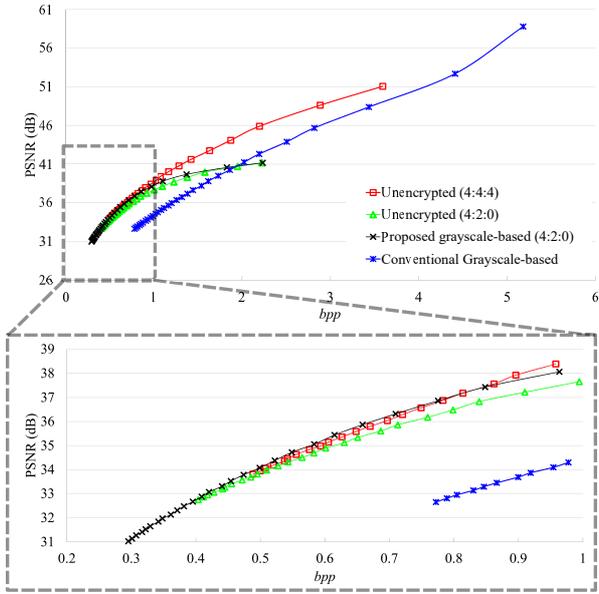}
\caption{R-D curves of JPEG images}
\label{fig:psnr-rd}
\end{figure}

Figure\,\ref{fig:psnr-rd} shows that the proposed scheme
has almost the same compression performance as non-encrypted ones with
4:2:0 color sub-sampling and also outperforms those without any
color sub-sampling and the conventional one. The proposed encryption scheme
allows us to avoid the effect of color sub-sampling and also improve compression performance of JPEG
compression.

\begin{figure}[t]
\captionsetup[subfigure]{justification=centering}
\centering
\subfloat[Twitter]{\includegraphics[clip,
width=8.5cm]{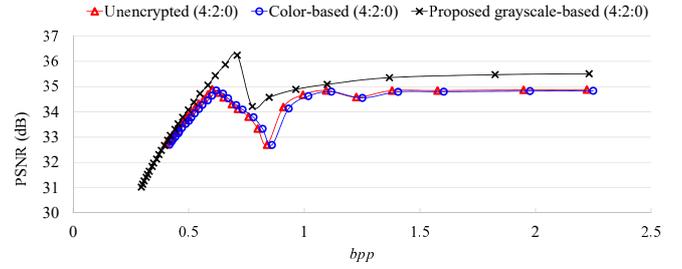}}
\\
\subfloat[Facebook]{\includegraphics[clip,
width=8.5cm]{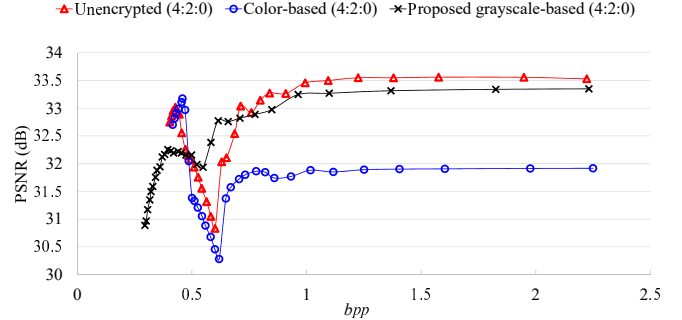}}
\caption{R-D curves of downloaded JPEG images}
\label{fig:sns-rd}
\end{figure}

\begin{figure}[t]
\captionsetup[subfigure]{justification=centering}
\centering
\subfloat[Color-based image encryption scheme
with $B_x=B_y=8$ and 4:2:0 sub-sampling (PSNR=26.21dB)]{\includegraphics[clip,
width=8.5cm]{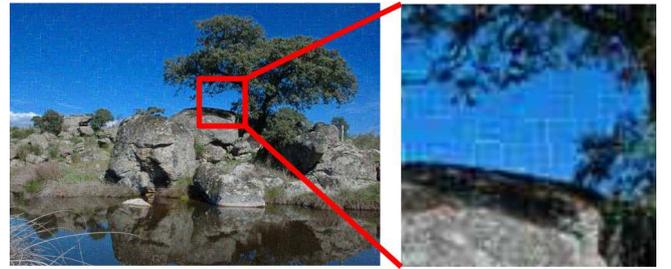}}
\\
\subfloat[Proposed scheme with $B_x=B_y=8$
(PSNR=31.85dB)]{\includegraphics[clip, width=8.5cm]{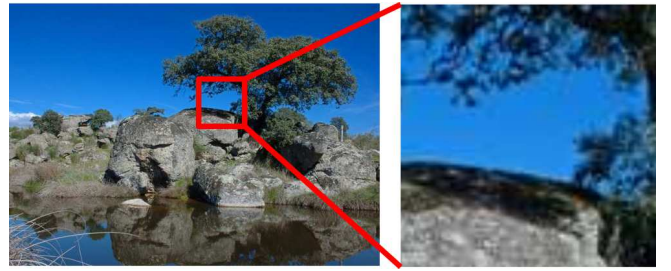}}
\caption{Example of decrypted images downloaded from Twitter}
\label{fig:dec_tw}
\end{figure}

\begin{figure}[t]
\captionsetup[subfigure]{justification=centering}
\centering
\subfloat[Color-based image encryption scheme with $B_x=B_y=8$ and 4:2:0
sub-sampling (PSNR=23.34dB)]{\includegraphics[clip,
width=8.5cm]{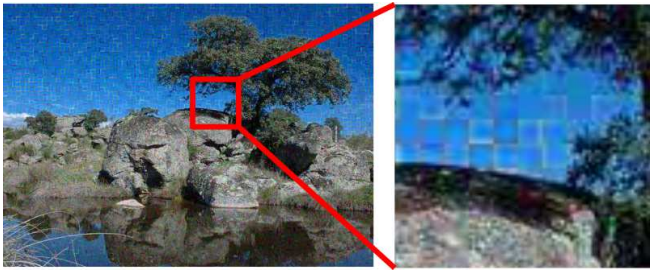}}
\\
\subfloat[Proposed scheme with $B_x=B_y=8$
(PSNR=29.96dB)]{\includegraphics[clip, width=8.5cm]{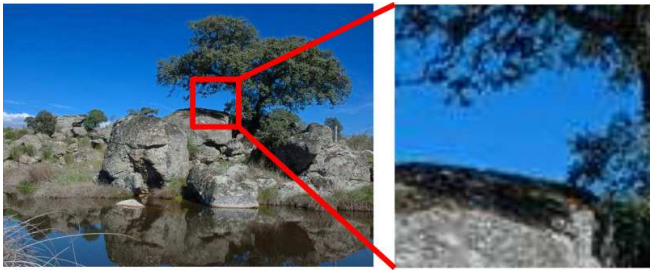}}
\caption{Example of decrypted images downloaded from Facebook}
\label{fig:dec_fb}
\end{figure}

\subsubsection{Compression Performance of Downloaded Images}
\label{sssec:performance_down}
According to image manipulation carried out by SNS
providers\cite{CHUMAN2017APSIPA}, an uploaded image is decoded and compressed
respectively based on their specifications when the uploaded image is satisfied
the conditions of SNS providers. 

In Fig.\,\ref{fig:sns-rd}(a) and (b), the performance of the color-based image
encryption and the proposed scheme was compared in terms of compression
performance of downloaded images from Twitter and Facebook, respectively. Note
that $B_x=B_y=16$ is used as a block size for the color-based image encryption
to avoid block distortion.

Twitter recompresses the uploaded JPEG images
with 4:2:0 color sub-sampling when $Q_{f_u}\geq 85$ to the new JPEG images with
4:2:0 sub-sampling ratio and $Q_{f_u}=85$. As shown in
Fig.\,\ref{fig:sns-rd}(a), the proposed scheme provided 
higher compression performance than those with the color-based image encryption
and non-encrypted ones.

In Facebook, every uploaded JPEG image is recompressed to the new JPEG
image with 4:2:0 sub-sampling ratio and $Q_{f_u}\in [71,85]$. The images
encrypted by the color-based encryption were heavily distorted by color
sub-sampling carried out by Facebook.
In comparison, the proposed one provided higher image quality compared
with the color-based image encryption. Moreover, as shown in
Fig.\,\ref{fig:sns-rd}(b), when $bpp>1$, PSNR values of the non-encrypted images
were higher than those with the proposed scheme approximately 0.2 dB. This is
because every grayscale JPEG image uploaded to Facebook is recompressed to the new grayscale JPEG image with $Q_{f_u}=71$ while Facebook recompresses JPEG color JPEG images with $Q_{f_u}\in [71,85]$.

Figures \,\ref{fig:dec_tw} and \,\ref{fig:dec_fb} show the example of decrypted
images downloaded from Twitter and Facebook, respectively. Since $B_x=B_y=8$,
the images encrypted by using color-based image encryption were strongly
distorted by the color sub-sampling carried out by the providers. In comparison,
the images encrypted by using the proposed scheme did not include any
distortion.

Considering color sub-sampling operation to the grayscale-based images
encryption does not affects the compression performance of JPEG images. The
results also proved that the proposed scheme can avoid the effects of color
sub-sampling carried out by SNS providers.

\section{Conclusion}

\label{sec:conclusion}
This paper considered color sub-sampling operation on the grayscale-based image
encryption for EtC systems. Firstly, the scenario and requirements of the image
encryption were described. Moreover, we proposed to generate the grayscale-based
image from the luminance and sub-sampled chrominance components. A lot of images
was compressed with 4:4:4 and 4:2:0 color sub-sampling ratio and decompressed to
evaluate the compression performance and the robustness against color
sub-sampling. The results proved that considering color sub-sampling operation to
the grayscale-based image encryption does not affect the compression performance
and also provides the robustness against color sub-sampling.

\bibliographystyle{ieeetran}
\begin{small}
\bibliography{refs}

\begin{thebibliography}{10}
\providecommand{\url}[1]{#1}
\csname url@samestyle\endcsname
\providecommand{\newblock}{\relax}
\providecommand{\bibinfo}[2]{#2}
\providecommand{\BIBentrySTDinterwordspacing}{\spaceskip=0pt\relax}
\providecommand{\BIBentryALTinterwordstretchfactor}{4}
\providecommand{\BIBentryALTinterwordspacing}{\spaceskip=\fontdimen2\font plus
\BIBentryALTinterwordstretchfactor\fontdimen3\font minus
  \fontdimen4\font\relax}
\providecommand{\BIBforeignlanguage}[2]{{%
\expandafter\ifx\csname l@#1\endcsname\relax
\typeout{** WARNING: IEEEtran.bst: No hyphenation pattern has been}%
\typeout{** loaded for the language `#1'. Using the pattern for}%
\typeout{** the default language instead.}%
\else
\language=\csname l@#1\endcsname
\fi
#2}}
\providecommand{\BIBdecl}{\relax}
\BIBdecl

\bibitem{huang2014survey}
C.~T. Huang, L.~Huang, Z.~Qin, H.~Yuan, L.~Zhou, V.~Varadharajan, and C.-C.~J.
  Kuo, ``Survey on securing data storage in the cloud,'' \emph{APSIPA
  Transactions on Signal and Information Processing}, vol. 3, e7, 2014.

\bibitem{lagendijk2013encrypted}
R.~Lagendijk, Z.~Erkin, and M.~Barni, ``Encrypted signal processing for privacy
  protection: Conveying the utility of homomorphic encryption and multiparty
  computation,'' \emph{IEEE Signal Processing Magazine}, vol.~30, no.~1, pp.
  82--105, 2013.

\bibitem{zhou2014designing}
J.~Zhou, X.~Liu, O.~C. Au, and Y.~Y. Tang, ``Designing an efficient image
  encryption-then-compression system via prediction error clustering and random
  permutation,'' \emph{IEEE transactions on information forensics and
  security}, vol.~9, no.~1, pp. 39--50, 2014.

\bibitem{Johnson_2004}
M.~Johnson, P.~Ishwar, V.~Prabhakaran, D.~Schonberg, and K.~Ramchandran, ``On
  compressing encrypted data,'' \emph{IEEE Transactions on Signal Processing},
  vol.~52, no.~10, pp. 2992--3006, 2004.

\bibitem{Liu_2010}
W.~Liu, W.~Zeng, L.~Dong, and Q.~Yao, ``Efficient compression of encrypted
  grayscale images,'' \emph{IEEE Transactions on Image Processing}, vol.~19,
  no.~4, pp. 1097--1102, 2010.

\bibitem{Hu_2014}
R.~Hu, X.~Li, and B.~Yang, ``A new lossy compression scheme for encrypted
  gray-scale images,'' in \emph{IEEE International Conference on Acoustics,
  Speech and Signal Processing (ICASSP)}, 2014, pp. 7387--7390.

\bibitem{watanabe2015encryption}
O.~Watanabe, A.~Uchida, T.~Fukuhara, and H.~Kiya, ``An
  encryption-then-compression system for jpeg 2000 standard,'' in \emph{IEEE
  International Conference on Acoustics, Speech and Signal Processing
  (ICASSP)}, 2015, pp. 1226--1230.

\bibitem{kurihara2015encryption}
K.~Kurihara, S.~Shiota, and H.~Kiya, ``An encryption-then-compression system
  for jpeg standard,'' in \emph{Picture Coding Symposium (PCS)}, 2015, pp.
  119--123.

\bibitem{KURIHARA2015}
K.~Kurihara, M.~Kikuchi, S.~Imaizumi, S.~Shiota, and H.~Kiya, ``An
  encryption-then-compression system for jpeg/motion jpeg standard,''
  \emph{IEICE Transactions on Fundamentals of Electronics, Communications and
  Computer Sciences}, vol.~98, no.~11, pp. 2238--2245, 2015.

\bibitem{Kuri_2017}
K.~Kurihara, S.~Imaizumi, S.~Shiota, and H.~Kiya, ``An
  encryption-then-compression system for lossless image compression
  standards,'' \emph{IEICE transactions on information and systems}, vol.
  E100-D, no.~1, pp. 52--56, 2017.

\bibitem{WARIT2018ICME}
W.~Sirichotedumrong, T.~Chuman, S.~Imaizumi, and H.~Kiya, ``Grayscale-based
  block scrambling image encryption for social network services,'' in
  \emph{IEEE International Conference on Multimedia and Expo (ICME)}, 2018.

\bibitem{WARIT2018ITCCSCC}
W.~Sirichotedumrong, T.~Chuman, and H.~Kiya, ``Compression performance of
  grayscale-based image encryption for encryption-then-compression systems,''
  in \emph{International Technical Conference on Circuits/Systems, Computers
  and Communications (ITC-CSCC)}, 2018, pp. 444--447.

\bibitem{CHUMAN2017APSIPA}
T.~Chuman, K.~Iida, and H.~Kiya, ``Image manipulation on social media for
  encryption-then-compression systems,'' in \emph{2017 Asia-Pacific Signal and
  Information Processing Association Annual Summit and Conference (APSIPA
  ASC)}, 2017, pp. 858--863.

\bibitem{Son_2016_CVPR}
K.~Son, D.~Moreno, J.~Hays, and D.~B. Cooper, ``Solving small-piece jigsaw
  puzzles by growing consensus,'' in \emph{IEEE Conference on Computer Vision
  and Pattern Recognition (CVPR)}, 2016, pp. 1193--1201.

\bibitem{Cho_2010_CVPR}
T.~Cho, S.~Avidan, and W.~Freeman, ``A probabilistic image jigsaw puzzle
  solver,'' in \emph{IEEE Conference on Computer Vision and Pattern Recognition
  (CVPR)}, 2010, pp. 183--190.

\bibitem{CHUMAN2017ICASSP}
T.~Chuman, K.~Kurihara, and H.~Kiya, ``On the security of block
  scrambling-based etc systems against jigsaw puzzle solver attacks,'' in
  \emph{IEEE International Conference on Acoustics, Speech and Signal
  Processing (ICASSP)}, 2017, pp. 2157--2161.

\bibitem{CHUMAN2017ICME}
------, ``Security evaluation for block scrambling-based etc systems against
  extended jigsaw puzzle solver attacks,'' in \emph{IEEE International
  Conference on Multimedia and Expo (ICME)}, 2017, pp. 229--234.

\bibitem{Gallagher_2012_CVPR}
A.~Gallagher, ``Jigsaw puzzles with pieces of unknown orientation,'' in
  \emph{IEEE Conference on Computer Vision and Pattern Recognition (CVPR)},
  2012, pp. 382--389.

\bibitem{JPEGLIB}
``Independent jpeg group,'' http://www.ijg.org/.

\bibitem{YANG2009}
E.~h.~Yang and L.~Wang, ``Joint optimization of run-length coding, huffman
  coding, and quantization table with complete baseline jpeg decoder
  compatibility,'' \emph{IEEE Transactions on Image Processing}, vol.~18,
  no.~1, pp. 63--74, Jan 2009.

\bibitem{ucid}
G.~Schaefer and M.~Stich, ``Ucid: An uncompressed color image database,'' vol.
  5307, pp. 472--480, 2004.

\end{thebibliography}
\end{small}
\end{document}